\documentclass[runningheads]{llncs}
\usepackage{graphicx}

\begin{document}
\title{On the Role of Intelligence and 
Business Wargaming in Developing Foresight}
\author{
Aline Werro\inst{1} \and 
Christian Nitzl\inst{1} \and 
Uwe M. Borghoff\inst{1,2}\thanks{Corresponding author: uwe.borghoff@unibw.de}
}
\authorrunning{A. Werro, C. Nitzl and U.M. Borghoff}
%
\institute{Center for Intelligence and Security Studies (CISS) \\ 
Universit{\"{a}}t der Bundeswehr M{\"{u}}nchen, Neubiberg, Germany
\and
Institute for Software Technology \\ 
Universit{\"{a}}t der Bundeswehr M{\"{u}}nchen, Neubiberg, Germany \\
\email{ciss@unibw.de}
}

\maketitle              

\begin{abstract}
Business wargaming is a central tool for developing sustaining strategies. It transfers the benefits of traditional wargaming to the business environment. However, building wargames that support the process of decision-making for strategy require respective intelligence. This paper investigates the role of intelligence in the process of developing strategic foresight. The focus is on how intelligence is developed and how it relates to business wargaming. The so-called intelligence cycle is the basis and reference of our investigation. 

The conceptual part of the paper combines the theoretical background from military, business as well as serious gaming. To elaborate on some of the lessons learned, we examine specific business wargames both drawn from the literature and conducted by us at the Center for Intelligence and Security Studies (CISS). It is shown that business wargaming can make a significant contribution to the transformation of data to intelligence by supporting the intelligence cycle in two crucial phases. Furthermore, it brings together business intelligence (BI) and competitive intelligence (CI) and it bridges the gap to a company's strategy by either testing or developing a new strategy. We were also able to confirm this finding based on the business wargame we conducted at a major semiconductor manufacturer.

\end{abstract}

\section{Background}
Over the centuries, the military has adapted and optimized its structures and concepts according to the experience gained in warfare. Similar to the military, companies are often operating in a quite hostile environment. This is particularly true for today's often saturated buyers' markets. In contrast to the military, business is not facing an enemy army but competition and instead of battling for territories they are battling for market shares and profits \cite{oriesek2008business}. Due to the common realities as well as the long experience and continuous optimization of military tools and concepts, the corporate world has become increasingly interested in them. Wargaming and intelligence belong to the later adaptions of military concepts into corporate settings.

Wargaming has a long tradition in military as a standard training tool to prepare officers for real situations and to support armed forces develop and test plans \cite{kurtz2003business}. Wargames accelerate learning and make it possible to train officers in peacetime and to test strategies without real casualties. While wargames are designed and dedicated to support strategic decision-making, they should not simply lead to a decision, but allow participants to learn from the decisions made \cite{perla1985wargaming,goztepe2014war}. 
Even though the business world adopted the fundamental principles of wargaming in the late 1950s \cite{hershkovitz2019wargame}, the added value to planning and decision-making is increasingly recognized in recent years. 
In a recent survey, Kowalik \cite{Kowalik19} showed that the primary function of business wargames is to improve decision making.

However, military and business wargames \cite{Schwarz2009} do not only share the basic elements of design but also their dependence on accurate intelligence. In order to build a meaningful wargame, the players do not only have to know about their own capabilities (business intelligence) but also their competitors and their environment (competitive intelligence). Instead of assessing the enemy army and studying the terrain, intelligence in the context of business wargaming is related to company strategies and general as well as industry-related trends. 

By the turn of the millennium, it was already clear that since the entry into a new phase of the information age \cite{BP1998}, an average private organization could access a quantity of data and information that was previously only available to intelligence agencies. In reference to a quote by Peter Senge -- {\em knowledge is the capacity for effective action} --, intelligence could also be defined as a means for effective strategy finding (e.g., \cite{huang2018research}). Small private intelligence departments have nowadays the capabilities to collect more data than what a state could do a decade ago. As relevant information became basically available to everyone at low cost, the competitive advantage is largely defined by how companies process the information and by the conclusions they draw from it \cite{soilen2017care}. This becomes even more important as artificial intelligence systems, which are an evolution of earlier knowledge-based systems, as well as knowledge discovery and data mining from big data, play a leading role in research and industrial deployment. 

Despite the abundance of information available, companies often struggle to turn the right information into intelligence and intelligence into strategy (e.g., \cite{cavallo2021competitive,herring1992role}). Therefore, this paper is investigating the role of intelligence in the process of developing strategic foresight. After specifying the terms business intelligence and competitive intelligence, the intelligence cycle and the associated challenges are emphasized. Common challenges that can arise when moving from intelligence to strategic foresight are then highlighted and business wargaming is presented as a possible aid to this task. 

The aim is to contribute to the debate of how intelligence can be integrated into the strategy of a company, which still lacks of research \cite{badr2006contribution,calof2017integration}. 
In order to support this debate, this paper falls back to military and governmental concepts of intelligence and connects it with the circumstances in the private sector.
We illustrate our conceptual work with examples from the literature as well as through a mentored business wargaming at a major semiconductor manufacturer within the 
Center for Intelligence and Security Studies (CISS).\footnote{https://www.unibw.de/ciss}

\section{Intelligence in Business}
Intelligence is implemented in corporate settings in order to improve a company´s competitiveness and strategic planning process \cite{guyton1962guide,saayman2008competitive}. Although intelligence has become an integral part of business management, no theoretical framework is prevailing. This is also reflected in the variety of different terms used in this context, such as business intelligence, competitive intelligence, corporate intelligence, market intelligence, strategic intelligence, etc. 

Another broadly used definition of intelligence in business describes intelligence as “a corporate capability to forecast change in time to do something about it. The capability involves foresight and insight, and is intended to identify impending change which may be positive, representing opportunity, or negative, representing threat” (\cite{breakspear2013new}, p.~678). However, in some cases where these concepts are applied, the term intelligence is omitted. For some companies, intelligence has a negative connotation (partly because some associate intelligence with secret state activity \cite{breakspear2013new,warner2002wanted} and partly because of the possible association to industrial espionage 
\cite{reinmoeller2016persistence,soilen2017care}, which is why the term is sometimes avoided. Instead, they are referred to as business analytics, competitive analysis, environmental scanning or subsumed under the term of management support systems.

In the context of this paper the discussion is focusing on the terms business intelligence (BI) and competitive intelligence (CI). Since some consider BI and CI as complements (e.g., \cite{saayman2008competitive}), while others understand them as synonyms 
(e.g., \cite{elena2011business,vedder2001study}, further specifications are needed. In the following, BI is considered internal intelligence about and within one´s firm, whereas CI is presumed external intelligence about the firm's competitors and environment, see, for instance, \cite{bose2008competitive,rouhani2012review}. This specification is particularly interesting with respect to business wargaming, where it is equally important to know the own skills as well as the competitors´.

When BI is referred to as an inward-looking view, it is considered a tool based on technologies, processes, and applications to collect and analyse data such as customer behaviour, inventory, manufacturing, etc.\ \cite{elena2011business,skyrius2021business}. 
The right IT infrastructure allows direct access to internal information which then is processed to support critical decision-making in real time \cite{boyer2010business}.

CI, on the other hand, monitors the external environment accordingly. The actionable intelligence should provide the firm with a competitive edge \cite{kahaner1997competitive}, by anticipating market developments \cite{bose2008competitive}. In the context of CI, there are different techniques to place the retrieved data in a useful context to support strategic decision-making, such as Porter´s five forces, the BCG matrix or the GE business screen matrix. These techniques allow to map the collected information in a way that addresses specific strategic decisions \cite{bose2008competitive,cavallo2021competitive}.

\section{Intelligence Cycle}
\begin{figure}[ht]
\centering\includegraphics[width=7cm]{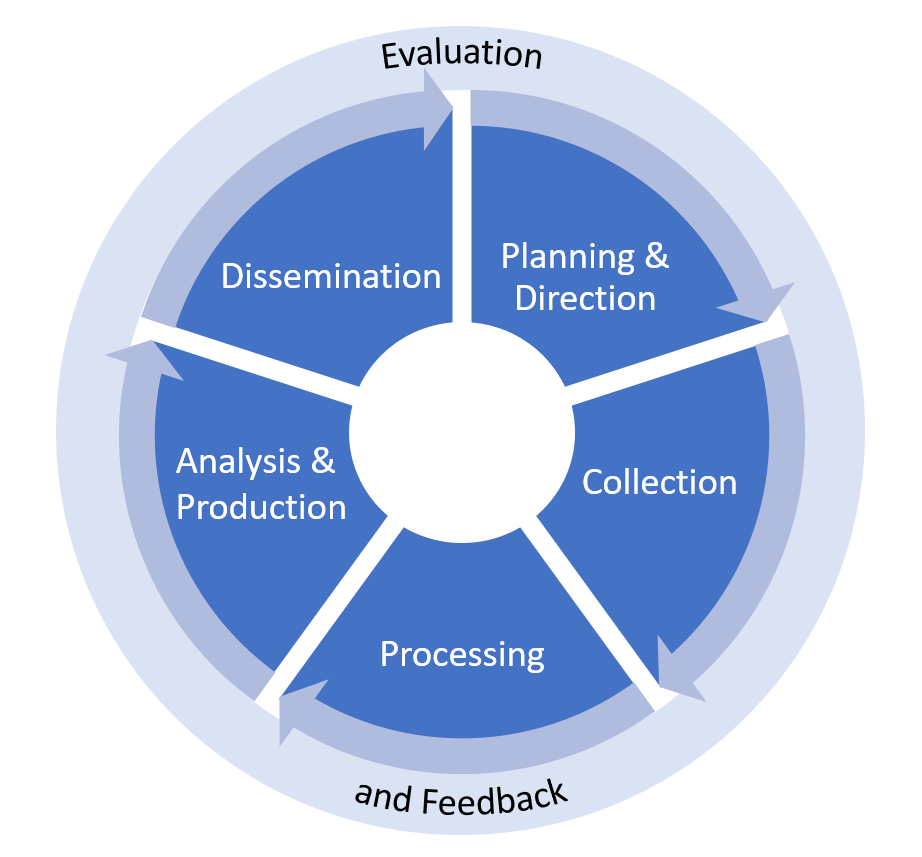}
\caption{Intelligence cycle.}\label{FIG1}
\end{figure}

The intelligence cycle is considered a systematic process designed to obtain intelligence from raw data \cite{aydin2015should}. Even though some authors may change the number or terminology of the elements, the intelligence cycle basically consists of five phases (see Figure~\ref{FIG1}):
\begin{enumerate}
\item planning and direction representing the beginning as well as the end of the cycle. It starts with the identification of the need for data and ends with an intelligence product that supports the decision makers and closes the cycle with new requirements. 
\item collection meaning the gathering of raw data. 
\item processing describing the conversion of the raw data into a suitable form.
\item analysis and production comprising the step from information to intelligence. The evaluated information is analysed and put into a respective context in order to obtain an assessment that allows further judgement.
\item dissemination addressing the distribution of the gained intelligence to the decision makers.
\end{enumerate} 
(Central Intelligence Agency, 1995 \cite{central1995factbook})

The need for a codified intelligence process for military staff set the foundation for the intelligence cycle \cite{nolan2015understanding,phythian2013understanding}. 
The constant evaluation and feedback loop has since been adopted to non-military areas of intelligence such as BI and CI. The circularity of the intelligence cycle indicates the direction of the process, but still allows for movement between the different phases. Despite ongoing arguments about the intelligence cycle being outdated 
(e.g., \cite{aydin2015should})
or not representative of how intelligence functions (e.g., \cite{Hulnick2006}), the cycle is still used to give crucial guidance.

Nolan (\cite{nolan2015understanding}, p.~114) claims that “information has become a firehose”. With the flood of available information, not only processing but above all planning and direction gained in importance. Without structured planning and direction, companies run the risk of the information overload becoming an obstacle rather than an opportunity. Consequently, an intelligence process is efficient and effective when it is focused on what is important to decision makers, rather than collecting as much data as possible \cite{cavallo2021competitive}. 
The decision makers are required to identify their intelligence needs, which are then translated into {\em key intelligence questions (KIQ)} \cite{Hulnick2006} or {\em key intelligence topics (KIT)} \cite{weiss2002brief} to give guidance to BI and CI. 

In practice, companies oftentimes do not only miss the guidance from planning and direction, but also a general strategy for their internal intelligence units. In 2011, Gilad \cite{gilad2011strategy} pointed out that over 90\% of all Fortune 500 companies use some form of CI, but when top executives were asked to recall one occasion CI has influenced their strategy or who their intelligence analyst is, they often go blank. 

\section{Internal Challenges on the Way from \\ Intelligence to Strategy}
More information does not necessarily lead to better decisions. Moreover, information is no longer the scarce resource it once was, especially since companies virtually all have the same access to it \cite{cavallo2021competitive}. As a consequence, the challenge {\em is not to deliver more information to anyone, at anytime, and from anywhere, but to provide the right information, at the right time, in the right place, in the right way to the right person} (\cite{fischer2012context}, p.~1). But even when BI and CI systems are able to turn information into actionable intelligence, a strategy still needs to be defined. Therefore, Herring stated that even {\em good intelligence, by itself, will not make a great strategy} (\cite{herring1992role}, p.~57).

Just as a strategy can only be as good as the information it derives from \cite{herring1992role}, the company must also have the required structures and culture in place to harness its intelligence. As emphasised by Søilen \cite{soilen2017care}, intelligence units are often confronted with job descriptions that encompass a wide range of tasks, such as sales or marketing. Furthermore, there is a lack of best practice in terms of where to position its intelligence unit within the company´s organigram. And lastly, executives can have a difficult relationship with intelligence analysts, seeing them as competitors rather than the support they are supposed to be. 

While the debate about the tight positioning (in terms of how close intelligence should be to policy) is not alien to intelligence services either (e.g., \cite{Hulnick2006}), it is, on the other hand, widely accepted that intelligence analysts are not the ones who make decisions. As intelligence analysts are usually specialised on a particular topic, decision makers are required to have a broader view. Just like a policy maker, a CEO has to take into account a wider range of considerations than an intelligence analyst. Therefore, decision makers rely not only on the obtained intelligence, but also on other considerations. 

As stated by Colebatch \cite{colebatch2006work}, some make choices and others help them to make the best choices. In order to be supportive, CI needs to be embedded in respective organizational structures. If the information is to be acted upon quickly, the mechanisms in place must ensure immediate dissemination of the information to the decision makers. One way to establish such mechanisms would be, for example, to include intelligence officers to weekly or monthly top management meetings, refer to \cite{calof2017integration}. 

Gilad \cite{gilad2011strategy} speaks of strategic intelligence (since CI could be mistaken as competitor watching) and warns that if companies do not find a way to overcome the internal competence issues, the price could be strategy without intelligence and intelligence without strategy. He also claims that one reason why some companies are reluctant to build intelligence capacity is that they find it difficult to define in advance what an intelligence analyst should be looking for.

While a lack in leadership can lead to BI and CI departments without a clear mission and a difficult standing within an organization, they can still run – just with less efficiency and effectivity. A wargame, on the other hand, forces the executives to actively engage. Before a wargame can take place, they need to determine what issues will be addressed. Without this minimum level of leadership, a wargame cannot come to fruition. The wargame itself then breaks down internal structures such as divisions, reporting structures, etc. and brings together people with different backgrounds and knowledge who are striving for the same goal.

\section{The Case for Business Wargaming}
The term business wargaming stands for a customised dynamic simulation that is specifically tailored to a company and the respective issues it is seeking answers to. The participants are divided into teams, each representing a different stakeholder. Typically, a business wargame lasts several rounds, with each round representing a specific period in the future. While the first period usually starts in the present, the time frame of the subsequent periods can span months, years or even decades, as required \cite{oriesek2008business}.

Wargaming is not about winning, but about the process of developing foresight and insight about possible actions and reactions in a dynamic and sometimes disruptive environment. Drivers of change from a competitive, political, sociocultural and technological environment can be included (see, for instance, Schwarz et al.\ \cite{schwarz2019combining}) as well as the findings from BI and CI. A wargame is thus preceded by an intensive preparation phase in which the required information is transformed into a design suitable for the objective at hand. However, the information serves not only to set up the teams, but also to create the models that are used to evaluate the teams´ decisions in the course of the game. The wargame is followed by a debriefing, which is crucial for the company's learning. It summarises the lessons learned and enables recommendations in the form of an action plan 
(e.g., Oriesek and Schwarz \cite{oriesek2008business} or Perla \cite{perla1990art}).

While one team represents its own company, the others slip into the roles of competitors, customers, etc. and are thus forced to change perspective. A manager knows the own strategy as well as the strengths and weaknesses of the company by heart. When taking up to role of a competitor in the context of a business wargame, he not only becomes an opponent to be taken seriously, but also gains a different perspective on his own company as well as on the role of the competitor. Gilad \cite{gilad2004early} therefore perceives business wargaming as the most effective managerial tool to assess the competitors' next moves.

Green and Armstrong \cite{green2011role} examined the effects of role-play on the forecasting accuracy in comparison to role-thinking and unaided judgement. The study presented experimental evidence that role-playing improved the forecasting performance to 60\% (while role thinking reached 32\% and unaided judgment 27\% accuracy). 

Taking a different perspective in role-playing has proven useful to overcome a variety of different cognitive biases. Hershkovitz \cite{hershkovitz2019wargame} further argues that wargaming can overcome organizational barriers. With knowledge -- that is necessary for strategic planning -- often split across different organizational functions, the design of wargames allows to bring together people with different perspectives from different functions with different interpretations of the operational environment. This also addresses the lack of collaborations where experts from different disciplines tend to create islands of knowledge that are otherwise not shared across the organizational channels. 
Albrecht et al.\ \cite{AlbrechtNB22} show how an integrated transdisciplinary approach can, for example, link the software development life cycle with the intelligence cycle. This leads to a forced sharing of knowledge between experts from the fields of law, psychology, political science, history, sociology, and computer science. 
Dobrovsky et al.\ \cite{DobrovskyBH16,DobrovskyBH19} address gamification for knowledge sharing and point out that balancing knowledge sharing (the serious part) and entertainment (the playful part) is still a challenge when adapting serious games to the needs of organizations.

In addition to the positive aspects of gamification, Perla and McGrady \cite{perla2011wargaming} highlight other benefits of serious gaming. Games engage the human brain on a different level than reading a report or watching a video. Participants are actively involved in experiencing a story, thus giving players responsibility for their actions and decisions. Strictly intellectual exercises such as scenario planning do not normally generate psychological or emotional stress.

A further central role for the future development of intelligence to strategy plays artificial intelligence. With AI-based wargaming, it becomes increasingly important to actually be able to prove the reliability of intelligence derived from AI models 
(e.g., \cite{SeidelB18,SeidelSB18}). 
In 2022, Davis and Bracken \cite{davis2022artificial} discuss decision support for wargaming, with and without AI, based on theory and exploratory work using simulation, history, and past wargaming. Although their work comes from political-military modelling, simulation, and wargaming of conflicts with nations, it could easily be applied to the business environment. Moy and Shekh \cite{moy2019application} investigate the process of automatic learning for wargaming with deep reinforcement learning. 
Seidel et al.\ \cite{SeidelSDB19} and Dobrovsky et al.\ \cite{DobrovskyBH16,DobrovskyBorghoff2017,DobrovskyWHHB17} also use deep reinforcement learning and show how artificial intelligence can improve processes in the intelligence cycle.

\section{Business Wargaming in the Context of the \\ Intelligence Cycle}
As Ginter and Rucks \cite{ginter1984can} point out, wargames are neither panaceas nor providers of solutions, but they offer a frameworks for addressing the {\em what if?} alternatives. Wargames bring together science and judgement and are therefore an important influence on strategic decisions. Furthermore, they improve the understanding of complex situations, facilitate the identification of problems, help to evaluate alternatives and their consequences \cite{kapper1981wargaming}. 
As Hernandez \cite{Hernandez15} shows, business wargames can even translate strategic budget decisions into terms of military effectiveness.

Most often, wargames are conducted either at the outset of the planning process or after a basic idea has been developed. At the outset, wargames can be a useful tool to convert the data and information that has been collected into actionable intelligence that can be incorporated in the subsequent planning of an organization. If a basic idea or a plan is the starting point of a wargame, robustness can be tested \cite{kurtz2003business}.

Consequently, wargames are usually conducted at the most crucial phases of the intelligence cycle. As the collection is determined by identifying potential sources and gathering them in a legal and ethical manner (e.g., \cite{bose2008competitive}), processing and dissemination are generally dictated by the structure and needs of an organization. However, planning and direction as well as analysis and production are the two phases where there is no standard procedure but where critical decisions are made.

We will now elaborate on and substantiate some lessons learned for these two crucial phases of the intelligence cycle through concretely conducted wargames: two that we took from the literature and one that we conducted ourselves at CISS.

\subsection{Wargaming at the planning and direction phase}
In the context of the intelligence cycle, wargames at the outset would facilitate the phase of planning and direction. Hence, a wargame at this phase can be of great value to determine which data – out of the masses available – should be collected and further processed. Without a coherent direction, an organization could end up collecting everything and yet nothing. At this phase, typical questions would be: {\em What information is needed? Why is it needed? When is it due?} (\cite{bose2008competitive}, p.~513).

\begin{figure}[ht]
\centering\includegraphics[width=8cm]{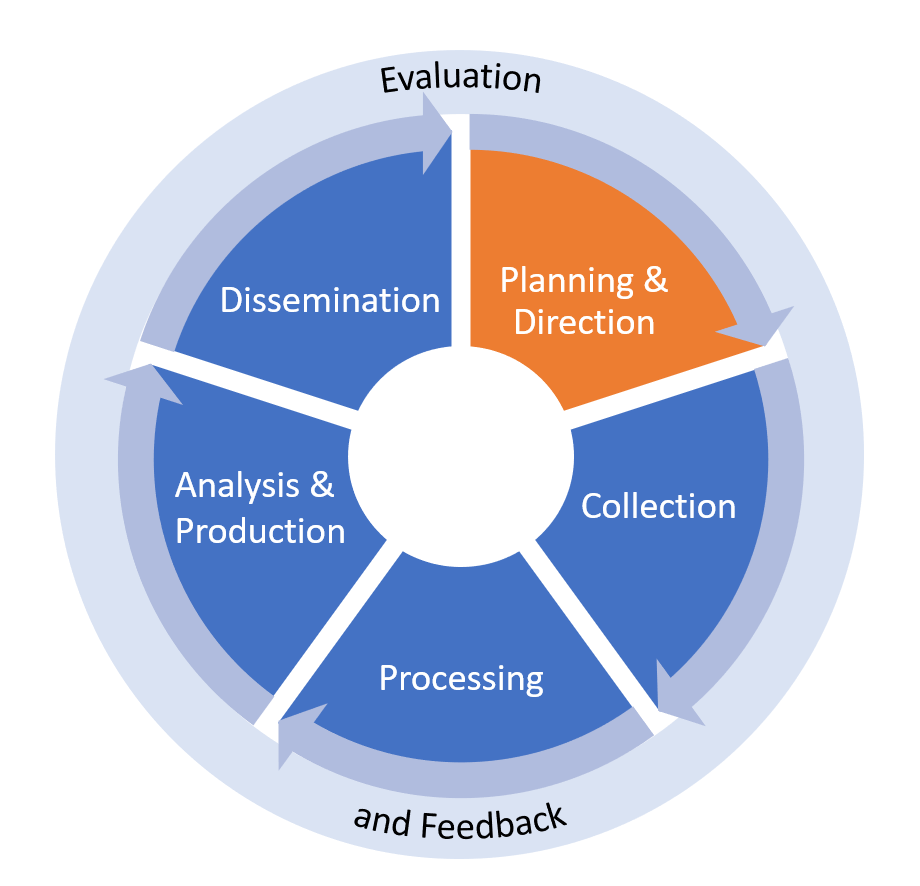}
\caption{Highlighting the planning and direction phase within the intelligence cycle for wargames at the outset.}\label{FIG2}
\end{figure}

As highlighted in Figure~\ref{FIG2}, planning and direction cannot pre-determine the kind of data and information that can be retrieved. In some cases, the anticipated information will not be available, while in other cases more information than expected can be retrieved. Despite a certain degree of unpredictability, an intelligible planning and direction phase is essential so that the intelligence analysts know what to aim for.

Kurtz \cite{kurtz2003business} gives an example of a business wargame conducted to support the planning and direction phase. The company was operating in a market where one competitor had a market share of about 60\%, while the company and the other smaller competitors had no more than 
8\%. The company wanted to develop a strategy to gain more market share. It already had numerous information about its competitors. However, the information was inconsistent and has not really been analysed to provide useful intelligence. The wargame was intended to help the management identifying {\em stronger and weaker players in the firm's markets and experiment with tactics that could improve its market share} (p.~13). The wargame led to the realisation that its {\em best strategy was to grow by taking business from other smaller players, without directly attacking the very powerful industry leader} (also p.~13). The company put this strategy into action and became a strong number two.

The role of the planning and direction phase in a wargaming is further explored below through a business wargaming at a major semiconductor manufacturer (Lead) accompanied by CISS.
The wargaming was conducted in 2021. In March 2021, a kick-off event was held in which the participants were introduced to the topic of business wargaming. In addition, instructions were given on how to put oneself in the role of the competitor. In May 2021, the actual business wargaming was conducted based on two rounds. In June 2021, the findings were presented to the division management.

Lead asked how the two main competitors (CompA and CompB) with the identical customer spectrum would react due to the strong increase in demand for discrete semiconductor components in the automotive industry. Lead is already the market leader in the market segment for modular but not for discrete semiconductor components. Especially due to the entry of new customers with a focus on discrete semiconductor devices, Lead expected the competitive situation to intensify under very limited production capacities. For data protection reasons, the here presented illustration was anonymised.

The wargaming was designed in such a way that two representatives in the first round presented Lead's planned strategy for the new upcoming customer requirements. For CompA and CompB, one team each was formed with four participants. These teams were formed of employees from the technical and commercial areas. This was to ensure that different facets of strategic decisions were comprehensively represented, such as the sales department being able to assess customer groups or the head of the engineering department being able to assess different production capabilities. All participants came from middle and senior management. A total of 13 participants actively took part in the business wargaming. Two participants represented the team Lead, four participants each represented a competitor, two persons from CISS and one person from Lead supervised the business wargaming.

The teams, which are to assume the role of the competitors, must be provided with comprehensive information about them. Lead already had numerous information about competitor CompA and CompB. This information included, for example, financial figures from the annual financial statements, company positioning from press releases or general information on market developments. However, it turned out that even though the available information was already very detailed, an important piece was missing in order to be able to get into the role of the competitor. This conclusion was based on the intelligence cycle, which states that in addition to the environmental conditions of a company, the intention of the management is also important in order to be able to predict future strategic behaviour. For this reason, in addition to the above-mentioned information, publicly available information on the management of CompA and CompB was collected in the planning and direction phase. This included demographic information, as well as information on educational background and professional career. This information can be found on a company's homepage or also in annual financial statements. In addition, press releases were searched for individual public statements made by management. This formed the background for the teams to be able to better understand the intentions of the respective competitor in addition to the capabilities.

As mentioned, the wargame we accompanied served to validate a new strategy of Lead and the reaction of competitors CompA and CompB to it. This constitutes an important part for the planning and direction phase. Since the company has figured out what it is aiming for, it became much easier to make sense of the information it already had and to fill in the gaps to develop intelligence.


\subsection{Wargaming at the analysis and production phase}
\begin{figure}[ht]
\centering\includegraphics[width=8cm]{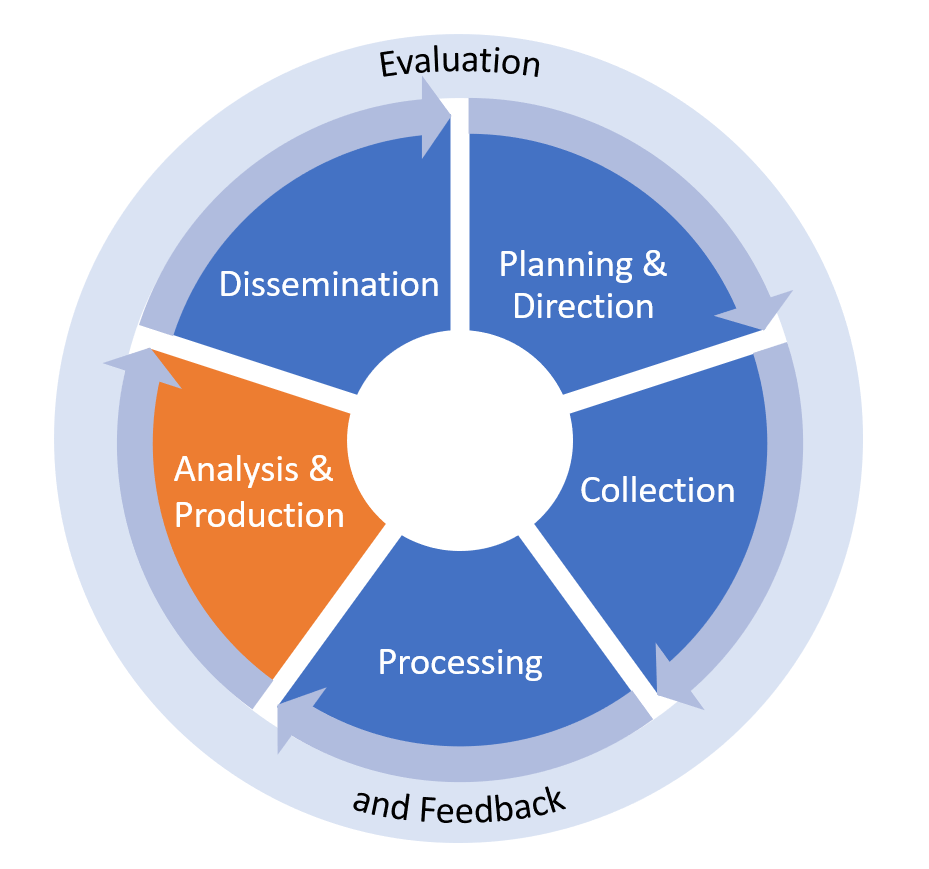}
\caption{Highlighting the analysis and production phase within the intelligence cycle for wargames testing a developed idea.}\label{FIG3}
\end{figure}

If a basic plan has already been developed, a wargame can also be held in the phase of analysis and production (see Figure~\ref{FIG3}). At this phase, the collected and processed information provides the basis for a wargame that addresses more advanced questions. Wargames are often designed to challenge a developed strategy and to show how the strategy might play out within the defined frame. Even when models quantify the potential revenue, market share, etc. in dependence of the respective strategy, the work is not done after the wargame is played out. The intelligence cycle is not driven by the knowns but by the unknowns. Therefore, one good question might lead to an answer, but also to several follow up questions that need to be answered accordingly. Consequently, the intelligence cycle is not completed after the dissemination of the analysis and production, but is redirected on the identified unknowns. 

Oriesek and Schwarz \cite{oriesek2008business} describe a case study concerning the changing environment for airlines due to the Open Skies regulation in the 1970s. Deregulation led to greater competition and falling prices, which led to the formation of alliances. A European airline aimed to test the viability of its current strategy, which was to establish a new alliance. The business wargame revealed that {\em the remaining candidates for this [new] alliance were in the main often poorly operated and weakly positioned} (p.~48). The main findings further suggested that {\em any existence outside of a major alliance would be increasingly difficult} (p.~47) and that {\em the bargaining power of any airline wanting to join a major alliance would decrease the longer it hesitated} (p.~47).

In this example, the CEO decided to pursue his original strategy despite the outcome of the wargame. The new alliance eroded the cash base of the airline without delivering any benefits from the created network, and in the following the airline went bankrupt. As the authors of the case study point out, wargaming might not have prevented this outcome, but it provided a clear indication that the creation of a new alliance at this stage would prove extremely difficult. Consequently, the undertaking should have at least been reconsidered. In the context of the intelligence cycle this would have meant to rerun the cycle given the new indications. Even if a wargame could not put off the CEO from his original strategy, he could have taken the hint about the questionable quality of the remaining potential alliance partners. Hence, he could have initiated a new run of the intelligence cycle to identify the potential partners for a new alliance and calculate the potential benefits that an alliance with these airlines would create.

For a deeper understanding of the importance of this crucial phase in the intelligence cycle, we revisit the business wargaming with a semiconductor manufacturer that CISS accompanied.
The goal of the conducted wargaming was to check the reaction of CompA and CompB to the newly developed strategy under the newly upcoming market developments. From the reactions of CompA and CompB it aimed then to determine what this would mean for Lead's strategy. In addition to many individual aspects, such as how competitors might position themselves with regard to individual semiconductor products, a scenario that had not previously been considered, or had only been considered very unlikely, was brought into focus. It turns out that for one of the played competitors CompB it could be very attractive to merge with another competitor existing on the market. This alternative has so far received little attention from Lead's management. However, it would have meant a completely new starting situation for the strategy planned at that time, due to the strong increase of new production capacities at the competitor.

In the further course, it could not be observed by CISS how much the knowledge gained from the business wargaming was taken into account in the further strategy development of Lead. In the subsequent period, however, it could be observed that numerous mergers and acquisitions took place on the semiconductor market. If even wargaming might not have prevented certain outcomes it provided a clear indication that mergers and acquisitions would be much more realistic as expected by Lead. 
Consequently, as in the previous example, the intention of the originally planned strategy would have had to be at least reconsidered. In the context of the information cycle, this would have meant restarting the cycle in view of the new indications.
In this business wargaming accompanied by CISS it could also be shown wargaming as a very effective managerial tool to assess the competitors' next moves and it is highly appropriated to improve decision making.

Despite the clear recommendations and all the insights gained in a business wargame, they are time and resource intensive and therefore do not (for most managers) offer a solution for strategy development in general. In fact, business wargaming is usually not applied to ordinary challenges, but rather on extraordinary ones. In contrast to the military, wargaming in business is not perceived as a standard tool, but as a tool -- out of the CI toolbox -- that offers a solution when the standard tools cannot. Consequently, it often depends on whether managers are willing to conduct a wargame and whether they are convinced of the added value of this method. Although many of these professional wargames were and are flawed -- which also means that current games for developing foresight are similarly flawed -- 
Curry \cite{curry2020professional} shows that such games are still invaluable for learning; 
see also 
Watt and Smith \cite{watt2021based} for a universal design for learning principles in which many of the obvious flaws can be fixed.

\section{Conclusion}
\begin{quotation}
{\em Tell me and I will forget, show me and I might remember, involve me and I will understand.} (Confuzius)
\end{quotation}

While companies have more data to their hands than ever before, it has also become a challenge to make sense of it. BI and CI offer several tools that can support the process of transforming data into intelligence. However, even if companies have well-functioning BI and CI departments, this does not necessarily mean that the intelligence will get to the right people at the right time in order to support the strategic decision-making. Business wargaming -- as one of these tools -- can make a significant contribution to this process. In addition to its ability to support the intelligence cycle in two crucial phases, the benefits of wargaming surpass standard business tools.

On the one hand, every qualitative and quantitative information a company has about its own business as well as about the competitors, market, customers, etc. can likewise be incorporated in a business wargame. On the other hand, business wargaming can help to overcome organizational barriers and bring together people from different divisions who would otherwise not necessarily have an exchange. In addition, aspects of serious gaming and role-playing in particular are positive impacts of wargaming. We also believe that deep reinforcement learning with artificial intelligence should be used more frequently for cloud-based big data analytics, especially in the analysis and production phase of the intelligence cycle. In particular, it is to be expected for the future that cloud applications will be inevitable in the processing of big data. However, such a cloud will be dominated by a few providers.\footnote{The dominant players such as Amazon (AWS), Google (Google Cloud), Microsoft (Azure) and IBM (Cloud Paks) are setting the rules and define AI based cloud functionality. Consequently, it becomes increasingly important to actually be able to prove the reliability of intelligence derived from AI models, especially when cyber-physical systems and humans are the subject of AI decisions, see \cite{SeidelB18,SeidelSB18}.}

Although business wargaming is considered one tool among others with specific characteristics, it is by no means interchangeable with other tools. Unlike the other tools, business wargaming not only supports BI and CI, but relies heavily on them. Without the insights form BI and CI, business wargames are not feasible. Consequently, they can help to identify what is still missing in BI and CI or support the connection of the different tools and create a bigger picture. In this way, business wargames are extremely helpful in connecting the dots. They bring BI and CI together and bridge the gap to a company's strategy by either testing or developing a new strategy. Furthermore, it requires active engagement and thus prevents blind flying of intelligence units.

But aspects of wargaming are not limited to military or business either. This paper has already touched on role-playing and serious gaming. However, there are also various other disciplines that can be associated with business wargaming (such as educational or psychological aspects). Therefore, research on business wargaming should not hesitate to connect different disciplines and mutually benefit from the research conducted.

\section*{Declaration of Conflicting Interests}
The authors declared no potential conflicts of interest with respect to the research, authorship, and/or publication of this article.

\section*{Data availability statement and Funding}
The participants of our study did not give written consent for their data to be shared publicly, so due to the sensitive nature of the research supporting data is not available.

The authors received no financial support for the research, authorship, and/or
publication of this article.

\section*{About the Authors}
{\bf Aline Werro} studied International Affairs and Governance at the University of St. Gallen (HSG), Switzerland. After her studies, she worked in the Swiss Federal Administration and dealt with various security and policy issues. Subsequently, she worked at the Geneva Centre for the Democratic Control of Armed Forces (DCAF) in the area of Security Sector Governance / Reform. Since March 2019, Aline Brülisauer has been working as a researcher at CISS at the University of the Bundeswehr in Munich. Aline Werro holds a special interest in game theory. Her focus is on the application of game theoretical concepts to current challenges in business and security policy. 

\medskip
\noindent
{\bf Christian Nitzl} is head of research for Wargaming and Information Systems at the Center for Intelligence and Security Studies (CISS). His focus is on research issues such as the art and science of wargaming, public management, public accounting, and statistical analysis methods. According to Stanford list (Scopus), he is among the top 2\%, and according to ResearchGate, he is among the top 1\% most cited scholars worldwide. He has published 50+ articles in peer-reviewed scientific journals.

\medskip
\noindent
{\bf Uwe M. Borghoff} studied computer science and mathematics and received his diploma and doctorate from the Technical University of Munich (TUM), Germany. He habilitated in Computer Science at TUM in 1993. He worked as a researcher at TUM for seven years before joining Xerox Research Center Europe at the Grenoble Laboratory, France, in 1994. In 1998, he joined the University of the Bundeswehr Munich, where he is a full professor at the Institute for Software Technology. Since 2004 he is Vice President of the University. From 2017 to 2023, he was the founding director of CISS.

%
%
%

%

\end{document}